\documentclass[11pt]{article}
\usepackage{latexsym}

\input epsf

\newtheorem{theorem}{Theorem}

\newtheorem{lemma}{Lemma}

\begin{document}

\title{ \bf Boundary Value Problem for $r^2 \,d^2 f/dr^2 + f = f^3$ (III):
 Global Solution and Asymptotics}

\author{Chie  Bing Wang\thanks
  {\small \it Current address:Department of Mathematics, University of California,
  Davis, CA 95616.e-mail:cbwang@math.ucdavis.edu } \\
 \small \it  Department of Mathematics, University of Pittsburgh \\
 \it Pittsburgh, PA 15260.}

\date{}
\maketitle

\begin{abstract}
Based on the results in the previous papers that the boundary 
value problem $y'' - y' + y = y^3$, $y(0) = 0, y(\infty) =1$
with the condition $y(x) > 0$ for $0<x<\infty$
has a unique solution $y^*(x)$, and
$a^*= y^{*^{'}}(0)$ satisfies $0<a^*<1/4$, in this paper we show that
$y'' - y' + y = y^3, \,\, -\infty < x < 0$, with the initial conditions
 $ y(0) = 0,     y'(0) = a^*$ has a unique solution by using
functional analysis method. So we get a globally well defined bounded
function $y^*(x)$,$-\infty < x < +\infty$. 
The asymptotics of $y^*(x)$ as $x \to - \infty$ and as $ x \to +\infty$ 
are obtained, and the connection formulas for the parameters
in the asymptotics and the numerical simulations are also given.
Then by the properties of $y^*(x)$,
 the solution to the boundary value problem 
$ r^2 f'' + f = f^3, f(0)= 0, f(\infty)=1$ is well described by the
asymptotics and the connection formulas.
\end{abstract}



\markboth{C. B. Wang}{ Monopole in the Pure SU(2) Gauge Theory}

\setcounter{equation}{0}
\section{Introduction}
  In this paper, we use the results obtained in 
\cite{wang1}  \cite{wang2} to study the 
 boundary value problem
    \begin{eqnarray}
  & &    r^2 f'' + f = f^3, \hspace{1cm} 0<r<\infty,    \label{eq1.1}   \\
 & & f(r)  \to 0,  \,  {\rm as \,\,} r \to 0,  \label{eq1.2}  \\
 & & f(\infty)=1.  \label{eq1.3}
    \end{eqnarray}
If we make a transformation
$r= e^x,   f(r) = y(x)$, the equation is changed to 
$y'' - y' + y=y^3$.

   The results in \cite{wang1}  \cite{wang2} can be briefly summarized 
as follows. 
The following boundary value problem 
   \begin{eqnarray*}
   & &  y'' - y' + y=y^3, \,\, 0<x< \infty,    \\
(P^+)& &y(0) = 0, y(\infty) = 1,    \\
   & &  y(x) > 0, \,\,\, 0<x< \infty,  
   \end{eqnarray*}
has a unique solution $y^*(x)$ for $0<x <\infty$.
 The formula of 
\begin{equation}
a^*= y^*{^{'}}(0), \label{addeq3.32}
\end{equation}
 is obtained, and 
\begin{equation}
0<a^* <1/4.  \label{addeq3.33}
\end{equation}
In this paper, we show that the initial value problem
   \begin{eqnarray*}
  & & y'' - y' + y = y^3, \,\, -\infty < x < 0,   \\
 (P^-) & &    y(0) = 0,      \\
   & &  y'(0) = a^*,    
   \end{eqnarray*}
has a unique solution by discussing the corresponding integral equation
on a Banach space. Then we get a global solution $y^*(x)$ of $y'' - y' + y=y^3$
 for $-\infty < x < +\infty$ and the asymptotics
as $ x \to -\infty$ and as $ x \to +\infty$.
And any bounded solution $y(x)$ of  this equation satisfying
$y(-\infty) = 0, y(+\infty) =1$ is expressed as
$y(x) = y^*(x - \tau)$, where $\tau$ is the largest zero of $y(x)$.
Therefore we have solved the boundary value problem 
(\ref{eq1.1})  (\ref{eq1.2}) (\ref{eq1.3}).

 This paper is organized as follows.
In Sect.~2, $y^*(x)$ is extended to the negative axis by
using contraction mapping theorem. In Sect.~3 we give 
numerical approximations for some important values 
including  $a^*$,  which are
used to represent asymptotics of the  solution. 
In Sect.~4, the asymptotic expressions and connection
formulas are given.

\setcounter{equation}{0}
\section{Global Solution}
   In \cite{wang1}, we have proved problem $(P^+ )$ has a unique
solution $y^*(x)=y(x,a^*)$. In this section we want to extend
the solution to the negative axis. Consider the following
problem
   \begin{eqnarray}
  & & y'' - y' + y = y^3, \,\, -\infty < x < 0, \label{eq6.1}  \\
 (P^-) & &    y(0) = 0,   \label{eq6.2}   \\
   & &  y'(0) = a^*.    \label{eq6.3} 
   \end{eqnarray}
Let
   \[ t=-x, \hspace{1cm} u(t) = e^{-x/2} y(x).  \]
Then  problem $(P^-)$ is equivalent to the
following integral equation
   \begin{equation}
   u(t)=- {2 \over \sqrt{3}} a^* \sin {\sqrt{3} \over 2} t
     + {2 \over \sqrt{3}} \int_0^t e^{-s}
      \sin{\sqrt{3} \over 2} (t-s) u^3(s) \, ds,  \label{eq6.10}
   \end{equation}
where $t \ge 0$.
Define
    \begin{equation}
    X = \{ f(t) | f {\rm \,\,is\,\, continuous\,\, on\,\,} [0,\infty),
        |f| \le {1 \over 2}, t \in [0, \infty) \},
               \label{eq6.15} 
    \end{equation}
    \begin{equation} 
    d(f,g) = \sup_{t \in [0, \infty)} |f(t) - g(t) |,
               \label{eq6.16}
    \end{equation}
for $f,g \in X$.
It is not difficult  to prove the following lemma.
\begin{lemma}     \label{Lemma6.1}
   (X,d) is a complete metric space.
\end{lemma}

Define
   \begin{equation}
  T(u)(t) =- {2 \over \sqrt{3}} a^* \sin {\sqrt{3} \over 2} t
     + {2 \over \sqrt{3}} \int_0^t e^{-s}
      \sin{\sqrt{3} \over 2} (t-s) u^3(s) \, ds,  \label{eq6.20}
   \end{equation}
for $u \in X$.

\begin{theorem}     \label{Theorem6.2}
$T$ has precisely one fixed point $u^*$ in $X$.
\end{theorem}

\noindent {\it Proof.}
   Let us first prove $T(u) \in X$ when   $u \in X$. By
(\ref{eq6.20}), $T(u)(t)$ is continuous on $[0, \infty)$.
By (\ref{addeq3.33}),
    \[ a^* < { 1 \over 4}, \]
which implies that  if $|u| \le 1/2$,
   \[  |T(u)(t)| \le {2 \over \sqrt{3}} a^*
             + {2 \over \sqrt{3}} 
            \left( { 1 \over 2} \right)^3
            < {\sqrt{3} \over 4} < {1 \over 2}.  \]
So $T(u) \in X$.

   Next we show $T$ is a contraction on $X$. In fact,
if $u_1, u_2 \in X$, there is
  \begin{eqnarray*}
  |T(u_1) - T(u_2)| &=&
   \left| {2 \over \sqrt{3}} \int_0^t
   e^{-s} \sin{\sqrt{3} \over 2} (t-s)
   (u_1 - u_2) (u_1^2 +u_1 u_2+u_2^2)  \, ds \right|  \\
   &\le& {3 \over 2 \sqrt{3}}
      \sup_{t \in[0,\infty)} |u_1(t)-u_2(t)| \,\,  \\
   &\le& {\sqrt{3} \over 2} d(u_1, u_2),
   \end{eqnarray*}
which implies
   \[
  d( T(u_1), T(u_2) ) \le {\sqrt{3} \over 2} d(u_1, u_2).
    \]
So $T$ is a contraction. By Banach fixed point theorem,
this theorem is proved.   \,\,\,\, $\Box$

\begin{theorem}    \label{Theorem6.3}
There is a unique solution $y^*$ to the problem
   \begin{eqnarray}
   & &  y'' - y' + y=y^3, \,\,\,\, -\infty<x< \infty,  \label{eq6.30} \\
 (P)  & &  y(0) = 0, \,\,\,\, y(\infty) = 1,  \label{eq6.31}  \\
   & &  y(x) > 0, \,\,\,\, 0<x< \infty.  \label{eq6.32}
   \end{eqnarray}
And $y^*$ has infinitely many zeros $x_n(n=0,1,2,...)$
  \[ -\infty<\dots<x_n<\dots<x_1<x_0=0,  \]
also
   \[ y(-\infty) = 0, \, |y^*(x)| \le {1 \over 2},  \]
for $x \in ( -\infty, 0]$.
\end{theorem}

\noindent {\it Proof.}
The Theorem 2 in \cite{wang1}  and Theorem 1  imply that problem (P) has a unique
solution $y^*$ . By (\ref{eq6.10}), we see that $y^*$ has
infinitely many zeros, and 
$y(-\infty) = 0, \, |y^*(x)| = |u^*(-x) e^{x \over 2}| \le {1 \over 2}$,
for $x \in ( -\infty, 0]$. \,\,\,\, $\Box$

\begin{theorem}      \label{Theorem6.4}
For any integer $n \ge 0$, there is a unique solution
$y^{(n)}$ to the following problem
   \begin{eqnarray*}
   & &  y'' - y' + y=y^3, 0 <x< \infty,    \\
   & &  y(0) = 0, y(\infty) = 1,   \\
   & & {\rm y \,\, has \,\, precisely \,\, n\,\, zeros\,\, in}\ (0,\infty).
   \end{eqnarray*}
\end{theorem}

\noindent {\it Proof.}
   By Theorem 2 , it is easy to check that
   \[ y^{(n)}(x) = y^* (x+x_n) \]
is a solution to this problem, where $x_n$ is given in Theorem 2.

    Now suppose there is another solution $\bar{y}^{(n)}$
to this problem. Let
  $  0= \bar{x}_0 < \bar{x}_1 < \bar{x}_2 < \dots < \bar{x}_n  $
be the zeros of $\bar{y}^{(n)}$ in $[0, \infty)$.
Then
  $ y(s) = \bar{y}^{(n)} ( s+ \bar{x}_n ),  $
satisfies (\ref{eq6.30}),(\ref{eq6.31}),(\ref{eq6.32}) with $x=s$. 
By Theorem 2  , we have
     $  y(s) = y^*(s),\,\, -\infty < s < \infty,  $
or
    $ \bar{y}^{(n)}(x) = y^* (x- \bar{x}_n).  $
Then $\bar{x}_k- \bar{x}_n (k=0,1,...,n) $ are the zeros
of $y^*$,
   $ y^*(\bar{x}_k - \bar{x}_n) = \bar{y}^{(n)}(\bar{x}_k) =0, $
satisfying
   $ - \bar{x}_n = \bar{x}_0 - \bar{x}_n < \bar{x}_1 - \bar{x}_n <
   \cdots < \bar{x}_k - \bar{x}_n < \cdots 
   < \bar{x}_n - \bar{x}_n = 0.  $
Because $0,\bar{x}_k(k=1, \dots, n)$ are the consecutive zeros
of $\bar{y}^{(n)}$, 
$\bar{x}_k- \bar{x}_n (k=0,1,...,n) $ are the consecutive zeros
of $y^*$. So
we see that
    $ x_n = - \bar{x}_n.  $
Thus
   $ \bar{y}^{(n)} (x) = y^*(x-\bar{x}_n)
      = y^*(x + x_n) = y^{(n)}(x).   $
So the theorem is proved.   \,\,\,\, $\Box$

\setcounter{equation}{0}
\section{Numerical Results}
   In this section, we present  numerical computation results
for some important values related to $y^*$. First of all, as seen
in the previous sections, the most important number is
$a^*$, which defines $y^*$. In the next section, we investigate
the asymptotics of $y^*(x)$ as $x$ approaches to $-\infty$
and as $x$ approaches to $\infty$. The following values
will be used to represent the asymptotics,

   \begin{eqnarray}
    b^* &=& \int_{-\infty}^0 e^{- {s \over 2}}
          \cos {\sqrt{3} \over 2} s \,(y^*(s))^3 \, ds,
                     \label{eq7.1}  \\
    c^* &=& \int_{-\infty}^0 e^{- {s \over 2}}
          \sin {\sqrt{3} \over 2} s \,(y^*(s))^3 \, ds,
                     \label{eq7.2}  \\
    d^* &=& \int_0^{\infty} e^s
     \left( (1-y^* (s))^2 -{1\over3}(1-y^* (s))^3 \right) \, ds.  
                         \label{eq7.3}
   \end{eqnarray}

{\bf \large (1) \,\,$a^*$ }

In Theorem 2 of \cite{wang1}, we obtained that
\[
\left( {1\over 2} +
    2 \sum_{n=1}^{N_1} {b_n \over n+1} \right)^{1\over2}
    < a^* <  - \sum_{n=1}^{N_2} b_n,
\]
for any $N_1 \ge 2$, and $N_2 \ge 1$.
If we choose large numbers  $N_1=170, N_2=170$, the { \it Mathematica }
shows the following result
   \begin{equation}
      0.16871221576 \cdots < a^* < 0.16871221594 \cdots .  \label{eq7.10}
   \end{equation}

\begin{figure}[ht]

\vskip 2cm
\vspace{-1.25cm}
\epsfxsize=300pt
\epsfysize=230pt 
 \centerline{\epsfbox{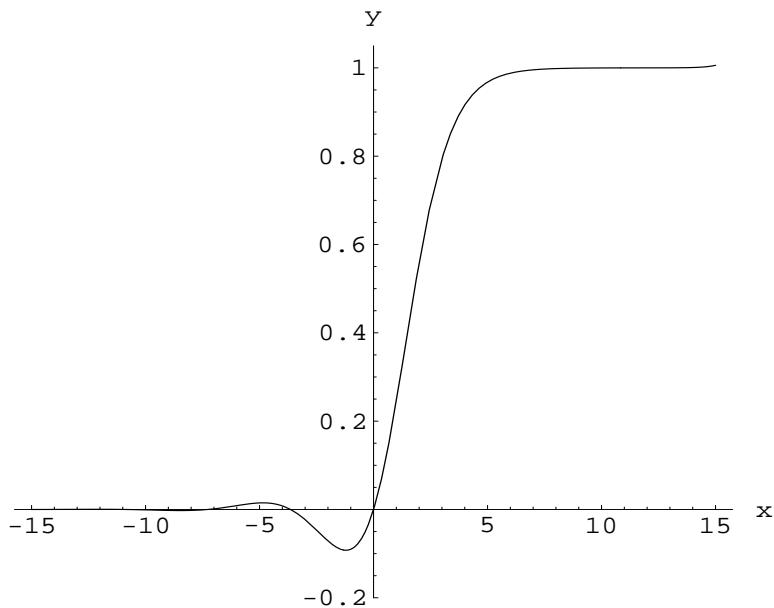}}
\vspace{1.25cm} 

\caption{ Graph of $y^* (x)$( solution of (P) ). The value of
 $a^* = {y^*}' (0)$ is  chosen  0.168712 $\cdots$. }
 \label{fig1} 
\end{figure}  

\begin{figure}[ht]
\vskip 2cm
\vspace{-1.25cm}
\epsfxsize=300pt
\epsfysize=230pt 
 \centerline{\epsfbox{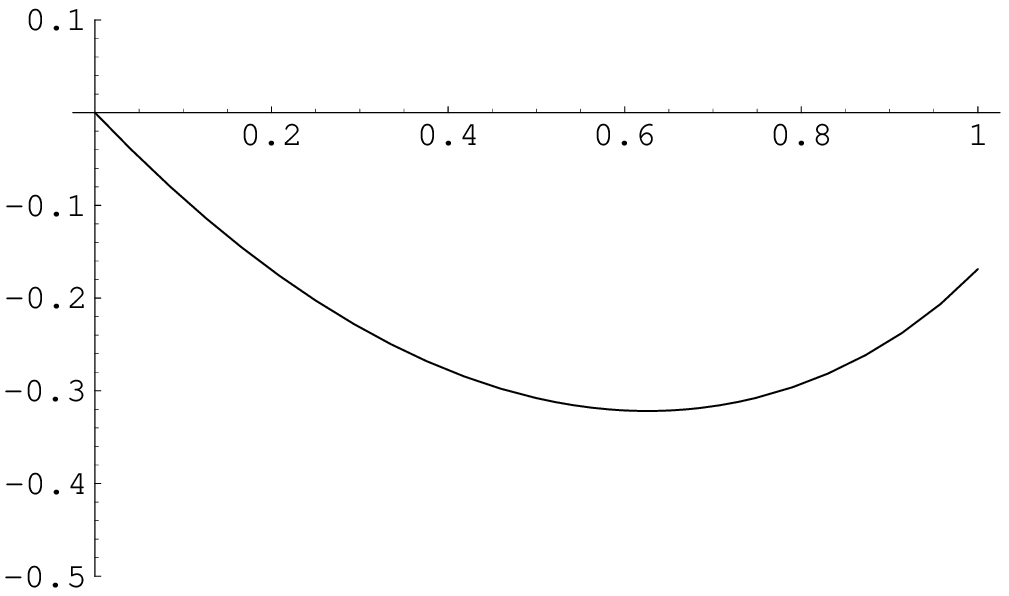}}
\vspace{1.25cm} 

\caption{ Graph of $P(z)$ as solution to the equation
  $ P P' - P = z(z-1)(z-2), z \in [0,1]$ with $P(z_0) =- z_0$,
  where $z_0 = 10^{-40}$. }
 \label{fig2} 
\end{figure}  

{\bf  \large (2) \,\,$y^*(x)$ and $P(z)$}

    In last section, we have proved that $y^*(x)$ is the unique solution to
problem (P). And it was also shown
 that $y^*(x)$ is  the solution of
the following initial value problem
    \begin{eqnarray}
     & &  y'' - y' + y=y^3, \,\,\,\, -\infty<x< \infty,  \label{eq7.11} \\
     & &  y(0) = 0, \,\,\,\, y'(0) = a^*.  \label{eq7.12}
    \end{eqnarray}  
 So we can approach the function
$y^*(x)$ by numerically solving the initial value problem with properly
approximate value of $a^*$ given above. The 
{ \it Mathematica} gives the  graph of $y^*(x)$. 
 See Figure 1. 

    In section 4 and 5, we discussed the function $P(z)$ which is
the solution to equation $ P P' - P = z(z-1)(z-2), z \in [0,1]$ 
with $P(z)  \sim - z$ as $ z \to 0$.
 As mentioned in \cite{wang2},
$P(z)$ gives the relation between ${y^*}'(x)$ and $y^*(x)$ for 
$x > 0$, which is
    \[ {y^*}'(x) = - P(1- y^*(x)).  \]
The graph of $P(z)$ is  computed also by 
{\it Mathematica}, and given in Figure 2. The numerical computation
for $P(z)$ shows that $P(1) \approx -0.1687 \cdots$, which matches
the result (\ref{eq7.10}).

{\bf \large (3) \,\,$b^*$, \, $c^*$, and  $d^*$}

   The values of $\,b^*, \, c^*, \, d^*$ will be used to represent
the asymptotics of $y^*(x)$ as $x \to \pm \infty$, 
and the asymptotics of $f(r)$
as $r \to 0$ and $r \to \infty$ in the next section.  We do not
have further analytic properties for these three numbers since we
have not found the analytic representation of $y^*$. Computer
shows the approximate values of these three  numbers 
based on the numerical computation for $y^*(x)$,
   \begin{eqnarray}
    b^* &\approx&  \int_{-30}^0 e^{- {s \over 2}}
          \cos {\sqrt{3} \over 2} s \, (y^*(s))^3 \, ds
        \approx - 0.0005497 \cdots,       \label{eq7.20}   \\
    c^* &\approx& \int_{-30}^0 e^{- {s \over 2}}
          \sin {\sqrt{3} \over 2} s \,(y^*(s))^3 \, ds
        \approx 0.001939 \cdots,       \label{eq7.21}   \\
    d^* &\approx& \int_0^{15} e^s 
       \left( (1-y^*(s))^2 - {1\over 3} (1-y^*(s))^3 \right) \,ds
      \approx  4.1728 \cdots.        \label{eq7.22}
   \end{eqnarray}

\setcounter{equation}{0}
\section{Asymptotics and Connection Formulas}
   We have seen in Sect.~2 that the problem (P) has a unique 
solution $y^*(x)$. Now, Let us consider the asymptotics of
$y^*(x)$ as $x \to -\infty$, and $x \to \infty$.

\begin{theorem}      \label{Theorem8.1}

The function $y^*(x)$ has the following asymptotics

   (i) As $x \to - \infty$,
   \begin{equation}
   y^*(x) = A e^{x\over2}
            \sin \left( {\sqrt{3} \over 2} x + \phi \right)
              + O(e^{3x \over 2}),  \label{eq8.9}
   \end{equation}
with the connection formula
   \begin{eqnarray}
      A &=& {2 \over \sqrt{3}} \sqrt{(a^*-b^*)^2+ {c^*}^2 }
             \approx  0.196,
                    \label{eq8.10}  \\
      \phi &=& \arctan {c^* \over a^* - b^*}
             \approx 0.0115 \approx 0.00375 \pi .
                    \label{eq8.11}
   \end{eqnarray}

   (ii) As $x \to 0$, there is
      \begin{equation}
      y^*(x) \sim a^* x.   \label{eq8.20}
      \end{equation}

   (iii) As $x \to \infty$,
   \begin{equation}
   y^*(x) = 1 - B e^{-x} + O(e^{-2x}),  \label{eq8.29}
   \end{equation}
with the connection formula
    \begin{equation}
     B = { {2 + a^* \over 3} + d^*} \approx 4.90.    
    \end{equation}

\end{theorem}

\noindent {\it Proof.}
Let us first consider (\ref{eq8.9}). Recall that if we let
    \[ t= -x, u^*(t) = e^{- {x\over2}} y^*(x),  \]
then $u^*$ satisfies for $t >0$
    \[  {u^*}'' + {3 \over 4} u^* = e^{-t} {u^*}^3 (t),  \]
and
    \begin{eqnarray*}
    u^*(t) &=& -{2 \over {\sqrt{3}}} a^* \sin {\sqrt{3} \over 2} t
      + {2 \over {\sqrt{3}}} \int_0^t \, e^{-s}
    \sin {{\sqrt{3}} \over 2} (t-s) {u^*}^3 (s) \, ds   \\
    &=& -{2 \over {\sqrt{3}}} a^* \sin {{\sqrt{3}} \over 2} t
      + {2 \over {\sqrt{3}}} \int_0^{\infty} \, e^{-s}
    \sin {{\sqrt{3}} \over 2} (t-s) {u^*}^3 (s) \, ds 
       + O (e^{-t}),
    \end{eqnarray*}
as $t \to \infty$. Direct calculation shows that
     \begin{eqnarray*}
     y^*(x) &=& {2 \over {\sqrt{3}}} (a^*-b^*) e^{x \over 2}
                 \sin {{\sqrt{3}} \over 2} x
              +{2 \over {\sqrt{3}}} c^* e^{x \over 2}
                 \cos {{\sqrt{3}} \over 2} x
             + O( e^{{3x} \over 2} )  \\
          &=& A e^{x \over 2} 
            \sin \left( {{\sqrt{3}} \over 2} x + \phi \right)
              + O( e^{{3x} \over 2} ),
      \end{eqnarray*}
where $A$ and $\phi$ are given by (\ref{eq8.10}),(\ref{eq8.11}).

  The asymptotics (\ref{eq8.20}) is obvious. Finally let us
prove (\ref{eq8.29}). Let
    \[ y^*(x) = 1 - z(x).  \]
It's easy to check that $z(x)$ satisfies the 
differential equation
   \begin{equation}
   z'' - z' - 2 z = -3 z^2 + z^3,  \label{eq8.40}   
   \end{equation}
and integral equation
   \begin{eqnarray*}
    z(x) &=& {{1 - a^*} \over 3} e^{2x} +
             {{2 + a^*} \over 3} e^{-x} +    \\
         & &   {{e^{2x}} \over 3} \int_0^x \,
                e^{-2s} ( -3 z^2(s) + z^3(s) ) \, ds
           - {{e^{-x}} \over 3} \int_0^{x} \,
                e^{s} ( -3 z^2 (s) + z^3(s) ) \, ds,
   \end{eqnarray*}
since $z(0)=1, z'(0) = - a^*$.
Using equation ({\ref{eq8.40}) and Lemma 3.1 (iii) in \cite{wang1}, we have
   \begin{eqnarray*}
   & &  \int_0^{\infty} \,
                e^{-2s} ( -3 z^2(s) + z^3(s) ) \, ds   \\
   &=&  \int_0^{\infty} \,
          e^{-2s} ( z''(s) - z'(s) - 2 z(s) ) \, ds   \\
   &=& -z'(0) -  z(0) 
        +( (-2)^2 + (-2) -2) \int_0^{\infty} \, e^{-2s} z(s) \, ds \\
   &=& a^* - 1.
   \end{eqnarray*}
Thus
   \begin{eqnarray*}
     z(x)  &=& {{2 + a^*} \over 3} e^{-x} 
             - { {e^{-x}} \over 3} \int_0^{x} \,
                e^{s} ( -3 z^2 (s) + z^3(s) ) \, ds  
            + O ( e^{-2x} )   \\
         &=& \left( { {2 + a^*} \over 3} + d^* \right) e^{-x} 
              + O ( e^{-2x} ).
   \end{eqnarray*}  
So (\ref{eq8.29}}) holds.   \,\,\,\, $\Box$ 

\begin{theorem}      \label{Theorem8.2}
Any solution $y(x)$ to the problem
     \begin{eqnarray}
  & & y''-y'+y=y^3, \,\, -\infty <x<\infty, \label{eq8.51}  \\
  & & y(x) \to 0, {\rm \,\,as\,\,} x \to -\infty \label{eq8.52}  \\
  & & y(x) \to 1, {\rm \,\,as\,\,} x \to \infty \label{eq8.53}
     \end{eqnarray}
has the representation
   \begin{equation}
      y(x) = y^* (x - \tau),  \label{eq8.55}
   \end{equation}
where $\tau$ is the largest zero point of $y(x)$.
Hence the solution to (\ref{eq8.51}), (\ref{eq8.52}),(\ref{eq8.53})
is unique up to a transition of $x$.

\end{theorem}

\noindent {\it Proof.}
Let $y_1=y, y_2=y'$, and change equation  (\ref{eq8.51}) into the system
        \begin{eqnarray*}
        y_1' &=& y_2,   \\
        y_2' &=& y_2 - y_1 +y_1^3.
        \end{eqnarray*}
It is not hard to see that the point $(0,0)$ in the phase plane is
a stable focus of the system (see \cite{perko}).
If $y(x)$ is a solution to  (\ref{eq8.51}), (\ref{eq8.52}) and
(\ref{eq8.53}), then $y(x)$ has infinitely many zeros. Because
$y(\infty) =1$, the zeros of $y(x)$ have upper bound, and
the largest zero $\tau$ exists(finite). Let
    \[ y_1(x) = y(x + \tau).  \]
Then $y_1(x)$ solves (P). By Theorem 2 , $y_1(x)=y^*(x)$, and then
(\ref{eq8.55}) holds.   \,\,\,\, $\Box$

   The solution to problem (P) is clear now. Let us come back to
the original  equation (\ref{eq1.1}).

\begin{theorem}    \label{Theorem9.1}
Any solution to (\ref{eq1.1}), (\ref{eq1.2}) and (\ref{eq1.3})
has the representation
   \begin{equation}
    f(r) = y^* (\log {r \over r_0} ),  \label{eq9.1}
   \end{equation}
where $r_0$ is the largest zero of $f(r)$. 
With $a^*, \, b^*, \, c^*$ and $ \,d^*$ defined 
by (\ref{addeq3.32}),(\ref{eq7.1}), 
(\ref{eq7.2}) and (\ref{eq7.3}) respectively, 
$f(r)$ has the following asymptotics.

   (i) As $r \to 0$, 
    \begin{equation}
     f(r) = A \left( {r \over r_0} \right)^{1\over2}
            \sin \left( {\sqrt{3} \over 2} \log {r \over r_0}
                + \phi \right) + O(r^{3 \over 2}),
                   \label{eq9.2}
    \end{equation}
where again
   \begin{eqnarray}
      A &=& {2 \over \sqrt{3}} \sqrt{(a^*-b^*)^2+ {c^*}^2 }
             \approx  0.196,
                    \label{eq9.10}  \\
      \phi &=& \arctan {c^* \over a^* - b^*}
             \approx  0.0115 \approx 0.00375 \pi .
                    \label{eq9.11}
   \end{eqnarray}

   (ii) As $r \to r_0$,
    \begin{equation}
      f(r) \sim a^* \log {r \over r_0}, 
    \end{equation}

   (iii) As $r \to \infty$,
    \begin{equation}
      f(r) = 1 - {B r_0 \over r} + O ( { 1 \over r^2 }),
    \end{equation}
where
   \begin{equation}
     B =  {2 + a^* \over 3} + d^*  \approx 4.90.    
   \end{equation}

\end{theorem}

\noindent {\it Proof.}
By Theorem 4  and Theorem 5 , 
using $r=e^x, r_0 = e^{\tau}, \log {r \over r_0} =x - \tau,
f(r) = y(x)$, direct calculations show
the results.   \,\,\,\, $\Box$


\end{document}